# Electrostatic, Luminescent, and Paramagnetic Responses of Fresh BN Nanopowders Synthesized under Concentrated Light

**Lina Sartinska**

*Department of Physics and Technology of Photoelectronic and Magnetoactive Materials, Frantsevich Institute for Problems of Materials Science, NASU, Omeliana Pritsaka str., 3, Kyiv 03142, Ukraine, email: l.sartinska@ipms.kyiv.ua*

**Abstract**

This study explores the properties of nanopowders synthesized under high-temperature, non-equilibrium conditions in a high-flux optical furnace in a nitrogen flow. Boron powders served as the starting material, and the intense thermal gradients during synthesis led to incomplete chemical reactions. As a result, the surface of the resulting nanoparticles is covered with a thin layer of sassolite, followed by boron oxides, beneath which lies a boron nitride shell. The subsurface contains boron-rich nitride phases, while the core consists of elemental boron. For reference, commercial platelet-like h-BN powders from the "Chempur" company were also analyzed.

Initially, all synthesized nanopowders displayed pronounced electrostatic charging, photoluminescence (PL), and paramagnetic activity, attributable to high surface defect densities and unsaturated chemical bonds. However, after two years of exposure to ambient air, these nanopowders exhibited complete loss of electrostatic charging, absence of photoluminescence (PL), and disappearance of the characteristic single EPR resonance line. Similarly, commercial h-BN nanopowder from the "Chempur" company does not exhibit a single EPR resonance line too. FTIR analysis revealed progressive surface oxidation and hydroxylation of this powder, suggesting that atmospheric moisture and oxygen effectively passivated defect states. These findings underscore the critical role of surface chemistry in governing the electrostatic, optical, and magnetic behavior of BN-based nanomaterials and highlight the importance of defect stabilization for preserving functional properties over time.

## 1. Introduction

The wide-bandgap semiconductor hexagonal boron nitride (h-BN) being a structural analogy to graphite is proving useful in electronics due to its chemical inertness, high thermal conductivity and stability, and resistance to oxidation and corrosion. Its unique properties also render it widely applicable in ceramics, functional composites, advanced insulators, hydrogen storage, thermal management, luminescence devices, sensing, adsorption, and catalysis [1], [2], [3], [4]. Recent research has focused on engineering boron nitride nanostructures with tailored properties to meet the demands of next-generation technologies [4].

It's known that nanosized crystallite has a much higher surface-to-volume atom ratio and, consequently, the intensity of such a parameter as photoluminescence (PL) should be correspondingly higher [5]. However, no evidence of photoluminescence has been found for some nanopowders. This indicates that PL in nanocrystalline powders is not only the result of a size effect but also a surface effect [5]. Investigation of surface defects has shown that despite the considerable density of point defects, which can be vacancies, interstitials, and antisites, several types of defects energetically lie either within the bands or in close proximity to the bands, resulting in the formation of shallow traps. By gradually tuning the densities of shallow and deep traps, it was observed that shallow traps can be healed upon exposure to illumination in the ambient, while deep traps are not affected [6].

During the synthesis process, as a result of external influences, different defects can be formed on the surface of BN nanoparticles [7]. Usually, h-BN easily forms a large-size sheet structure during the synthesis process due to its crystal characteristics. Research of photoluminescence of boron nitride nanosheets exfoliated by the ball milling method has demonstrated that at room temperature, the nanosheets have strong deep ultraviolet light emission due to the stacking faults caused by the shear force during milling [8].

PL of h-BN has been studied with nanosecond UV laser irradiation (248 nm) below ablation threshold in different environments: under vacuum, nitrogen atmosphere, and in ambient air. Laser fluence and irradiation dose affect the PL spectra. The first observed bands are tentatively ascribed to single and multiple nitrogen vacancies obtained in nitrogen atmosphere. Irradiation can break B-N bonds producing boron enriched surface under vacuum. However, no elementary boron appears under irradiation of samples in ambient atmosphere. This effect is explained by the oxygen healing of radiation-induced surface defects. Formation of oxide layer prevents the B-N dissociation and preserves the bulk sample stoichiometry [9].

Using the full potential linearized augmented plane wave (FLAPW) method has shown that magnetic properties of hexagonal boron nitride (h-BN) are strongly dependent on the defect type (adatom or vacancy defect) existing in the h-BN monolayer and this finding may help reveal the origin of magnetism in the h-BN layer [10].

Although both nitrogen and boron vacancies can act as paramagnetic centers in h-BN [11], one-boron centers (OBC) and three-boron centers (TBC) are the most investigated paramagnetic centers [12]. The one-boron centers are observed in samples that underwent oxidation. Three-boron centers are found in samples with carbon impurities [13]. On the other hand, TBC centers can also be produced by high-energy electrons in h-BN in the absence of carbon impurities [14]. Neutron irradiation also induces an alternative type of paramagnetic defect [15] [16] [17].

Defects of h-BN use to result in oxygen attaching to the surface in an oxygen-rich environment [18]. For example, EPR signal can be removed after annealing for 15 min at 140 °C in a vacuum and correlates with the observed degradation of BN film properties. This result could be related to absorbed hydronium ions ($H_3O+$) between the planes in the hexagonal component of the films [19], or to other complex defects involving oxygen and hydrogen. Oxygen, boron, and nitrogen atoms can adsorb and form a triangle above the center of the B–N bond. This means that h-BN will inevitably attach oxygen atoms in an oxygen-rich environment [20].

Electrostatic interactions at the nanoscale which are highly versatile and tunable, arise from a complex interplay of particle size, shape, dielectric properties, and environmental conditions, and are fundamental to the engineering of functional nanomaterials [21]. Experimental and theoretical studies have shown that external electric fields can effectively modulate the electronic structure of boron nitride nanotubes and nanosheets [22], [23], [24]. Point defects in BN nanosheets introduce localized charges that reshape the electrostatic potential landscape and influence their reactivity, polarity, and electronic transitions [25]. Electrostatic interactions in h-BN increase nanofriction under applied bias in nanoscale devices [26]. The use of electrostatic forces also enables the alignment of boron nitride (BN) platelets through flocking within flexible polymer matrices [27]. Studies have demonstrated that electrostatic deformation (electrostrain) decreases adhesion and shear strength in suspended multilayer h-BN, suggesting a promising approach for minimizing wear in nano-/microelectromechanical systems (N/MEMS) [28]. Oxygen doping in boron nitride (BN) induces the formation of paramagnetic $OB_3$ centers, which enhance light absorption into the deep visible region. This phenomenon underscores a significant relationship between electrostatic surface states and band gap engineering [29].

Despite extensive research on boron nitride, many aspects of its intrinsic and extrinsic defect structures—particularly those governing surface-dependent optical and magnetic behavior—remain insufficiently understood. This is especially true for BN nanostructures synthesized under concentrated light. In this work, we investigate the electrostatic charging behavior, the photoluminescence (PL) and electron paramagnetic resonance (EPR) properties of BN nanopowders with varied structural and chemical characteristics. The study emphasizes the role of surface structure and environmental exposure in shaping the functional response of BN, and evaluates the potential of PL and EPR techniques as sensitive probes for surface state characterization.

## 2. Experimental

BN nanopowders were prepared from two partially amorphous nanopowders. Equiaxed β-rhombohedral boron nanopowder had a grain size of 0.05 μm. β-tetragonal boron powder with a grain size of 0.20 μm and stabilized by an α-$B_{49.94}C_{1.82}$ had an impurity of $B_2O_3$. h-BN of the "Chempur" company was chosen for compering. The phase composition of initial and resulting nanopowders was analyzed using X-ray diffraction (diffractometer "DRON-3.0", radiation of $K_{α-Cu}$).

Synthesis of BN was carried out in a quartz chamber in a xenon high-flux optical furnace with xenon sources of exposure in a flow of nitrogen [30] [31] [32]. A surface of a compacted sample of boron powders has been heated up in a focal zone of the optical furnace at the density of energy in the focal zone set-up ~ 0, 7 × $10^4$ kW/$m^2$. The time of the experiment was 15 minutes.

Spectra of prepared BN powders were obtained using Spectrometer Nicolet 6700 FTIR [20]. The resulting structures were examined by scanning electron microscopy Superprobe-733. The photoluminescence (PL) method characterized the optical properties of nano-BN at room temperature. The light source LED-NS375L-5RLO (375 nm, 6 mW) was used for the excitation of light emission [34]. The electron paramagnetic resonance (EPR) spectra were recorded on h-BN samples using spectrometer EPR ELEXIS 580 also at room temperature.

## 3. Results

Concentrated light heating of initial boron powders due to high temperatures and significant temperature gradients leads to incomplete reactions according to the results of the X-ray diffraction study. The phase composition of both powders besides BN contains B, $B_2O_3$, and $B_{18}N_2$ (Fig. 1). Being active, boron powder with a grain size of 0.05 μm contributed to the formation of large platelet-like particles which had a thickness of 0.001 μm and a mean grain size of 0.5 μm (Fig. 1 a). Stabilization by an α-$B_{49.94}C_{1.82}$ boron powder with a grain size of 0.20 μm results in the formation of equiaxed particles with a size of 0.1 μm, which contains also an additional amorphous phase and $B_{1.1}N_{0.9}$ (Fig. 1 b). Platelet-like nanoparticles of h-BN of the "Chempur" company with impurity of $B_2O_3$ had a thickness of 0.001 μm and a mean grain size of approximately 0.3 μm (Fig. 1c).

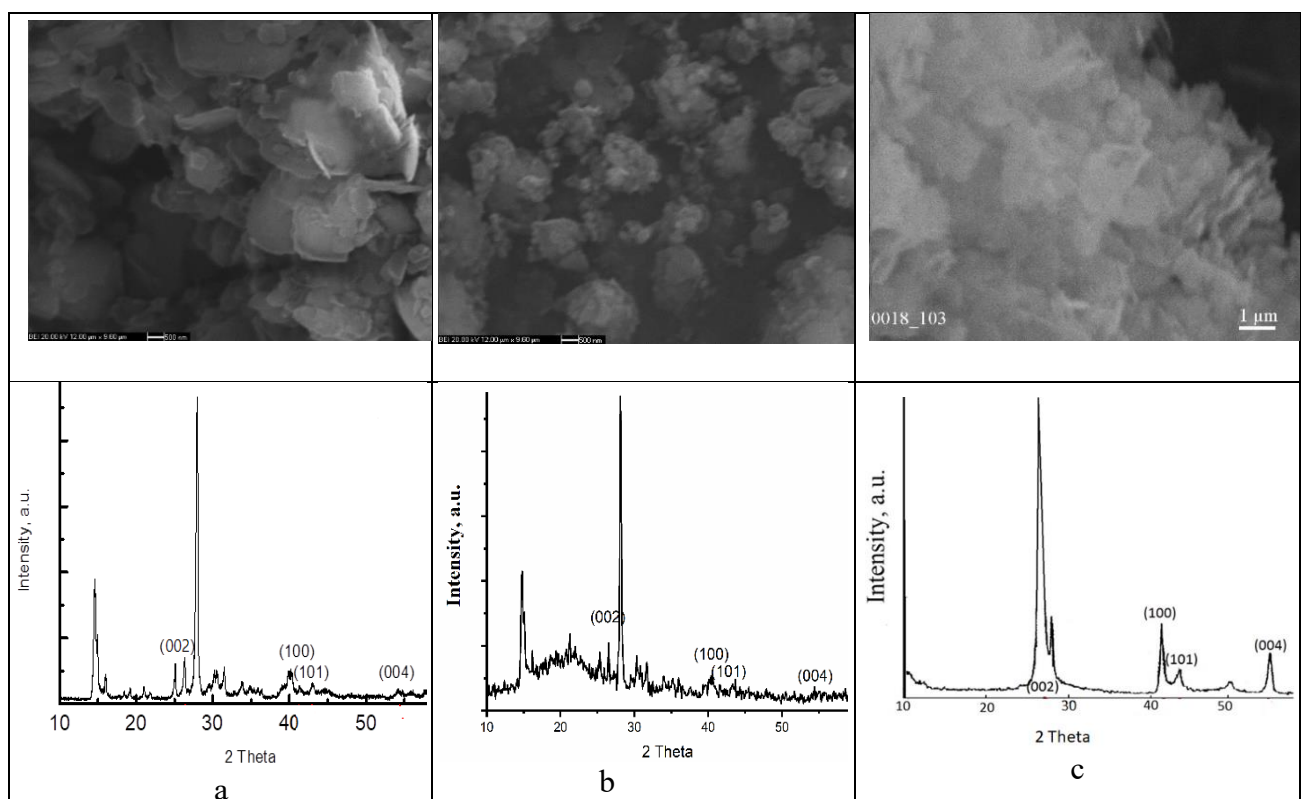

Fig. 1. SEM image and X-ray diffraction pattern of BN nanoparticles produced in a xenon high-flux optical furnace in a flow of nitrogen from boron of sizes: a - 0.05 μm and b - 0.20 μm. Nanopowder of h-BN of "Chempur" company was chosen for comparing (c).

All FTIR spectra of presented powders have a B-N stretching band at 1365 cm$^{-1}$ (h-BN) and an out-of-plane B-N-B bending band at 780 cm$^{-1}$ (Fig. 2). Spectra of prepared BN nanopowders are close to each other and differ from the h-BN of "Chempur" company. There are bands for B-O stretching in the region of 1200–1600 cm$^{-1}$, and oxygen-related functional groups B-O-N bonds form in the 900–1200 cm$^{-1}$ region due to adsorption from the environment on the surface of BN. The bands at 2260 cm$^{-1}$ and at 3195 cm$^{-1}$, which correspond to the stretching vibration of B–H bonds and N–H, respectively, are not typical for pure BN and also can indicate the surface adsorption of gases and post-synthesis exposure to air/moisture. This result confirms the XRD study and suggests a complex gradient or layered structure of the particles [35]. Based on the above, it's possible to conclude that the particle's surface is covered by a thin film of sassolite ($H_3BO_3$), followed by boron oxides ($B_2O_3$), beneath which boron nitride is located. Deeper layers consist of boron nitrides with an increased boron content ($B_{18}N_2$; $B_{1.1}N_{0.9}$), and at the very core, there is pure boron (Fig. 3).

It's known that the energy of the exciting light quanta ~3.3 eV ($\lambda_{exc}$ = 375 nm) is not sufficient for band-to-band excitation since the band gap of h-BN ranges from ~5 eV to ~6 eV. Defect-related levels inside the band gap of all BN make possible exciting by light quanta ~3.3 eV ($\lambda_{exc}$ = 375 nm) [34]. Complicated phase composition, presence of additional functional groups (Fig. 2), defects, and impurities introduced during synthesis of two prepared nanopowders result in a low intensity of photoluminescence spectra with numerous additional bands (Fig. 4 a). h-BN of the "Chempur" company demonstrates one band of higher intensity that is typical for pure BN (Fig. 4 b).

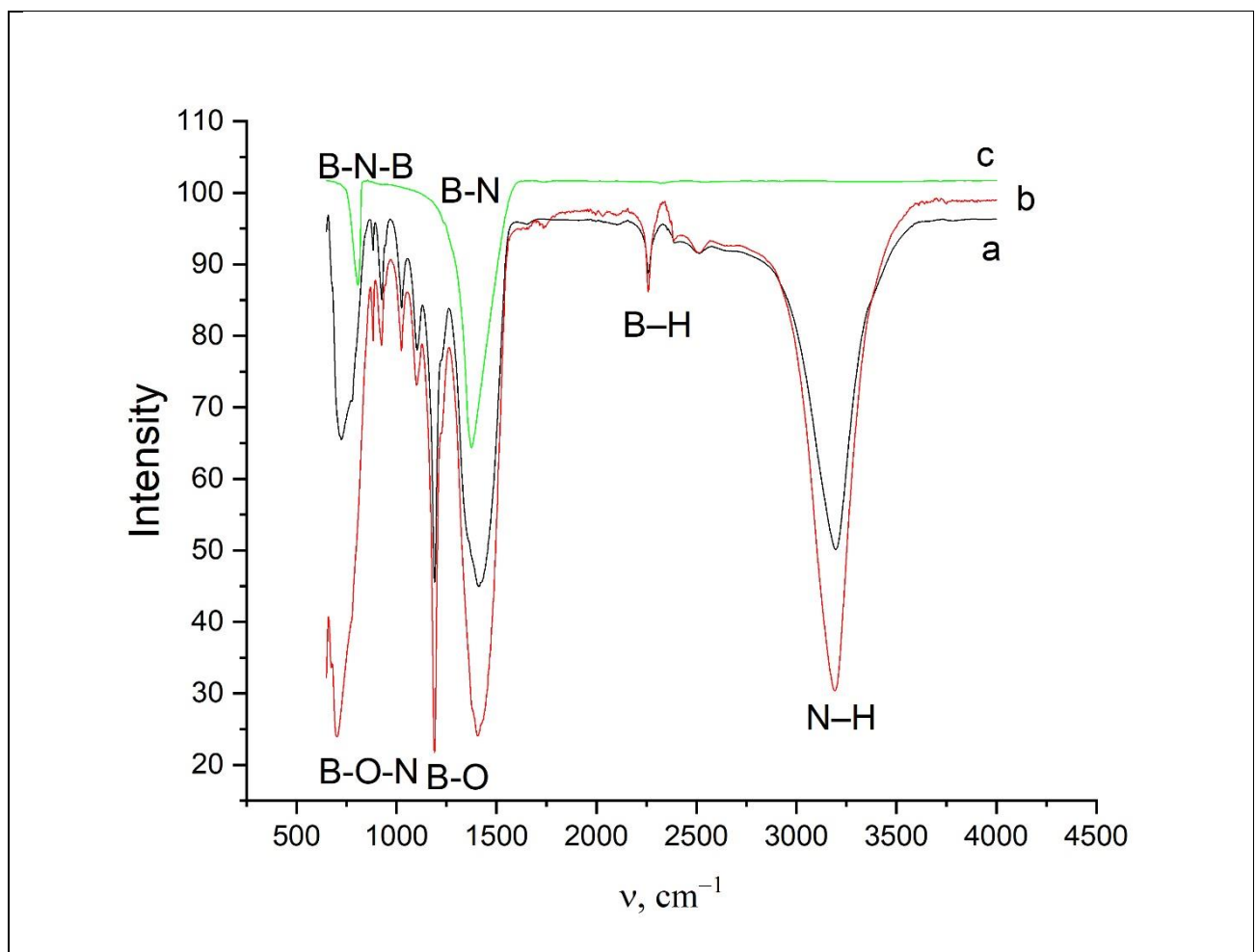

Fig. 2. FTIR spectra of BN nanoparticles produced from boron of sizes: 1 - 0.05 μm and 2 - 0.20 μm. Nanopowder of h-BN of "Chempur" company was chosen for comparing (3).

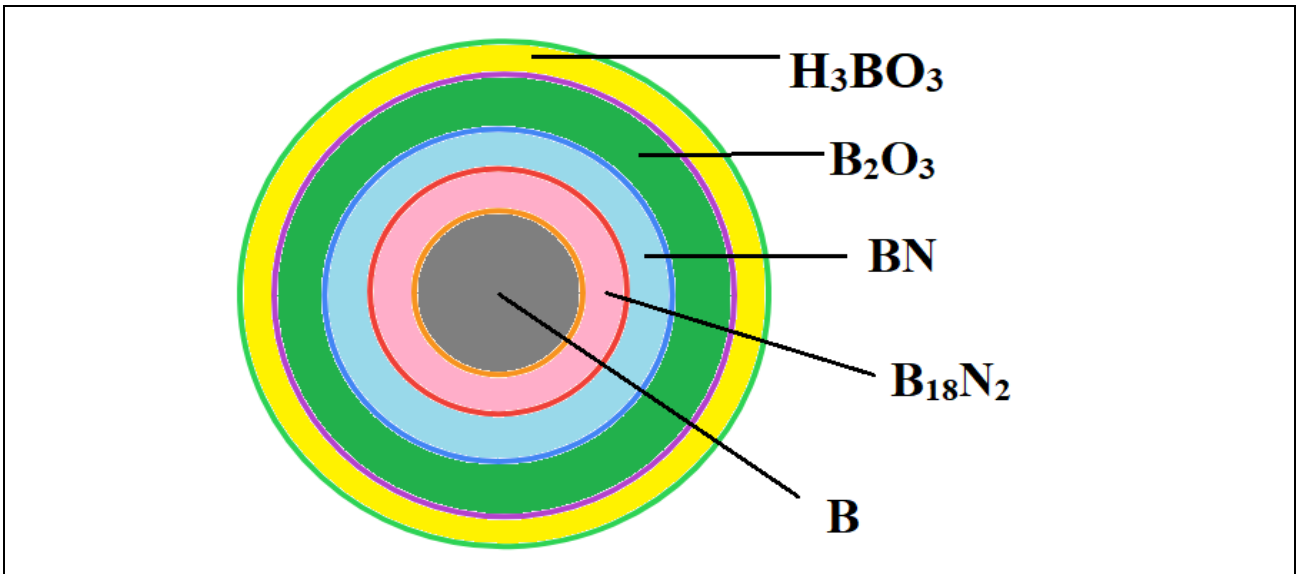

Fig. 3. Schematic representation of a layered or gradient structure in nanoparticles synthesized under concentrated light in an optical furnace in nitrogen flow, illustrating phases formation resulting from incomplete reaction.

The as-received BN nanopowders displayed pronounced electrostatic charging, allowing them to interact with external magnetic fields and spontaneously assemble into chain-like structures. This behavior is indicative of strong surface charge retention and interparticle polarization, which further supports the high defect density and non-equilibrium nature of the synthesized materials. The commercial h-BN nanopowder of "Chempur" company does not exhibit noticeable electrostatic behavior. Its high degree of crystallinity and low concentration of structural defects—as evidenced by X-ray diffraction (XRD) and Fourier-transform infrared (FTIR) spectroscopy—combined with a likely passivated surface, effectively suppress charge accumulation and interparticle polarization, thereby preventing the formation of chain-like or aggregated structures.

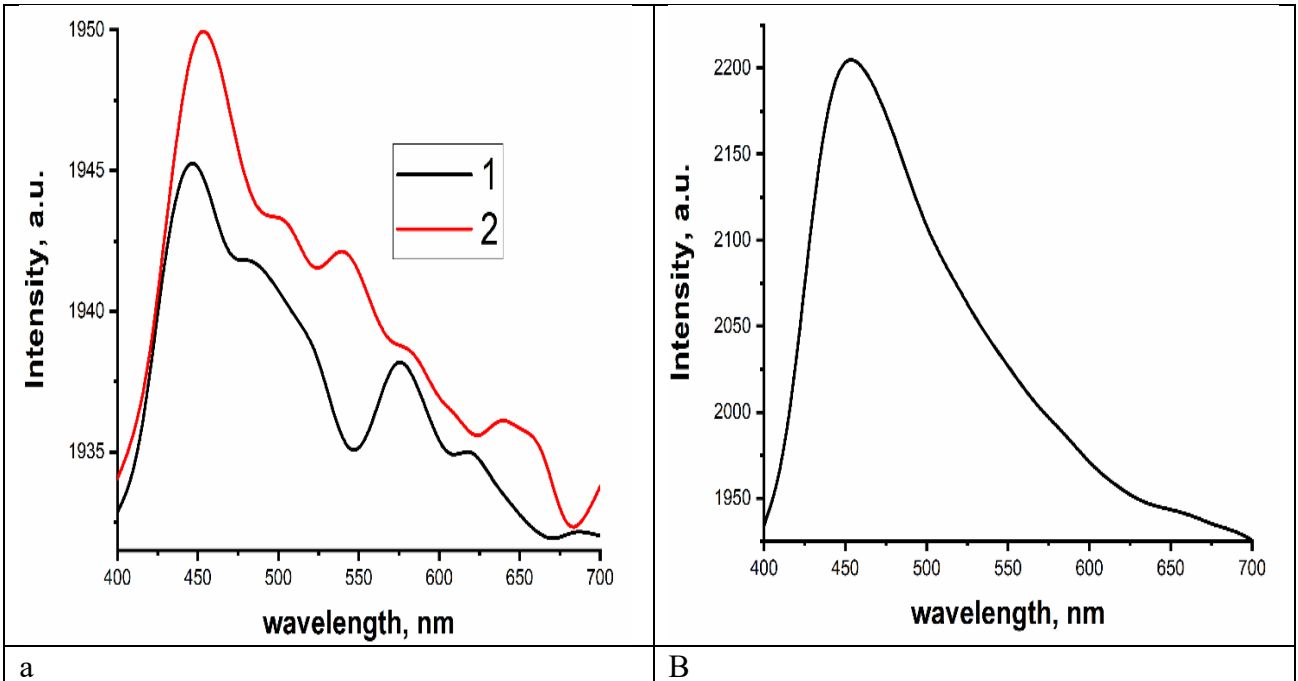

a | B

Fig. 4. Photoluminescence spectra at excitation wavelength 375 nm of BN nanoparticles produced from boron of sizes: 1 - 0.05 μm and 2 - 0.20 μm (a). Nanopowder of h-BN of "Chempur" company was chosen for comparing (b).

One of the key parameters that provide information about a substance's structure from EPR spectra is the g-factor, which measures the effective magnetic moment of an electron and determines the position of the absorption band in a magnetic field. The g-factor of a "free" electron is 2.0023.

The first reported defect of BN with g = 2.0023 consisting of 10 lines in EPR spectra was explained as a hyperfine interaction of an unpaired electron with three equidistant $^{11}$B atoms [36].

The EPR spectra of all investigated BN nanopowders do not exhibit hyperfine structure (Fig. 5). A single Gaussian resonance line for BN nanoparticles produced from boron of sizes: 0.05 μm, 0.20 μm and h-BN of "Chempur" company with g-values of 2.0058, 2.0047, and 2.0075, respectively, was found. The equiaxed structure of particles for BN nanopowder produced from boron powder with a particle size of 0.20 μm and stabilized by an α-$B_{49.94}C_{1.82}$ exhibit more isotropic magnetic properties compared to elongated, platelet-like or irregularly shaped particles, usually tend to localize unpaired electrons and to show a single, more uniform g-value due to reduced spin-orbit coupling and directional dependence. As a result, the g-value in equiaxed BN nanoparticles is closer to the free-electron g-factor. Furthermore, due to their equiaxed morphology, BN nanopowders synthesized from boron powder with a particle size of 0.20 μm and stabilized by α-$B_{49.94}C_{1.82}$ exhibit significantly lower EPR spectral intensity compared to other nanopowders produced using different synthesis methods.

The peak-to-peak line widths for all BN samples studied are 10, 33, and 3 G, respectively. Since lower g values often indicate weaker spin-orbit coupling, this can result in shorter spin relaxation times, leading to broader peak-to-peak line widths for BN nanopowder of equiaxed structure prepared from boron powder with a particle size of 0.2 μm (Fig. 1b). However, the equiaxed structure of BN nanopowders can also increase the EPR peak-to-peak linewidth due to higher defect density, leading to exchange broadening, and increased surface effects, modifying the local magnetic field. Thus, equiaxed BN nanopowders tend to exhibit broader EPR lines due to increased paramagnetic center interactions and shorter spin coherence times.

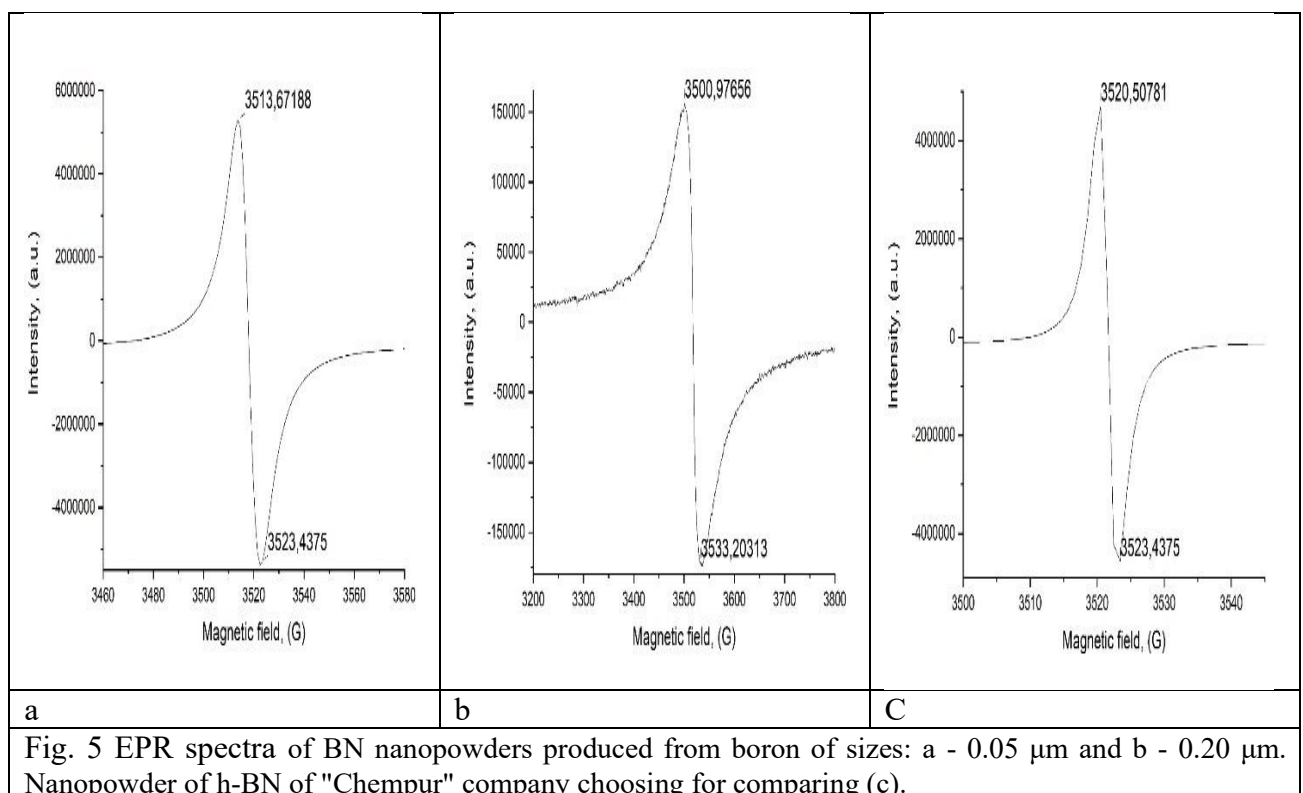

| a | b | C |

Fig. 5 EPR spectra of BN nanopowders produced from boron of sizes: a - 0.05 μm and b - 0.20 μm. Nanopowder of h-BN of "Chempur" company choosing for comparing (c).

After two years of exposure to ambient air, BN nanopowders synthesized under concentrated light exhibited complete loss of electrostatic charging, absence of photoluminescence (PL), and disappearance of the characteristic single EPR resonance line. Similarly, commercial h-BN nanopowder from the "Chempur" company does not exhibit a single EPR resonance line too. For the

BN powders synthesized in a xenon high-flux optical furnace under nitrogen flow, this degradation can be attributed to the loss of surface activity. Stabilization of the starting boron powder using α-$B_{49.94}C_{1.82}$ did not prevent the loss of the EPR signal.

## 4. Discussion

Freshly synthesized BN nanopowders initially exhibited strong UV photoluminescence, which is attributed to surface and near-surface defect states. Upon prolonged air exposure, consistent with surface healing and defect passivation mechanisms PL intensity significantly diminishes. These observations underscore the dominant role of surface states in governing the optical properties of BN nanomaterials. Intentional oxygen doping of h-BN, which introduces paramagnetic $OB_3$ centers, with g-factors ranging from 2.0030 to 2.0060, extends light absorption into the deep visible range and remains stable even after prolonged air exposure [29]. BN nanopowders, synthesized under non-equilibrium conditions in a xenon optical furnace, initially exhibit paramagnetic behavior with g-values in a similar range (2.0047–2.0075). However, these signals vanish entirely after ambient exposure, accompanied by the disappearance of photoluminescence and electrostatic activity. This behavior suggests that, unlike the intentionally introduced oxygen defects, our materials undergo spontaneous surface healing — likely through passivation by adsorbed oxygen and moisture, which neutralizes or removes shallow surface states responsible for EPR activity. Such discrepancy may point to fundamentally different types of paramagnetic centers: intrinsic, deep-level defects stabilized by substitutional oxygen, versus surface-localized or adsorbate-associated centers in our case. This suggests a strong link between surface electrostatic states and band gap modulation. Although BN nanopowders synthesized under concentrated light were not intentionally doped, the loss of PL and EPR activity following ambient exposure suggests a similar surface transformation mechanism.

The observed disappearance of EPR activity upon air exposure in BN nanopowders synthesized under concentrated light aligns with defect-healing mechanisms known from other studies. Paramagnetic centers such as boron or nitrogen vacancies are highly sensitive to the local surface environment. For instance, neutron-irradiated h-BN loses its EPR activity upon annealing due to oxygen-induced healing of vacancies [37]. Early studies have shown that EPR-active centers, such as F-centers (color centers) and carbon-related defects, are primarily localized at or near the surface, and their signal diminishes with surface relaxation or chemical adsorption [38]. In cubic BN, surface defects are also known to give rise to color centers and EPR signals, both of which are highly sensitive to moisture and other adsorbed species [39]. These results support conclusion that the degradation of EPR activity in BN nanopowders over time is driven by surface chemical transformation—including water and oxygen adsorption—and consequent passivation of defect sites. Therefore, our findings highlight a key distinction between engineered and ambient-induced surface modification of BN nanostructures. While controlled oxygen doping can stabilize and tailor the defect landscape for functional applications, uncontrolled post-synthesis oxidation may lead to the loss of functional properties, including paramagnetism and luminescence. Understanding and managing this surface chemistry is thus crucial for the reliable application of BN-based nanomaterials.

FTIR spectroscopy of the h-BN nanopowder "Chempur" company after two years of exposure to ambient air confirms these transformations. The FTIR spectra exhibit a noticeable broadening of the B-N stretching band at 1365 cm$^{-1}$, characteristic of h-BN, along with a shift in the out-of-plane B-N-B bending mode from 780 to 811 cm$^{-1}$ (Fig. 6). The broadening of the B-N stretching band can be attributed to increased phonon scattering, which arises from defects or local distortions in the h-BN lattice caused by exposure to air. This effect is primarily due to the adsorption of oxygen, water molecules, and other gases. Additionally, water vapor can lead to hydroxyl (-OH) functionalization at boron or nitrogen sites, further disrupting the local bonding environment and contributing to the broadening of vibrational modes. The shift in the out-of-plane B-N-B bending mode from 780 cm$^{-1}$ to 811 cm$^{-1}$ can be attributed to the formation of surface-bound species such as B-O, B-OH, or N-OH. These interactions alter the local bonding environment, increasing bond stiffness and consequently shifting the vibrational frequency to a higher wavenumber. Thus, our results align with and reinforce the emerging view that controlled surface chemistry in BN nanostructures can enable functional tuning of their electronic and photonic behavior.

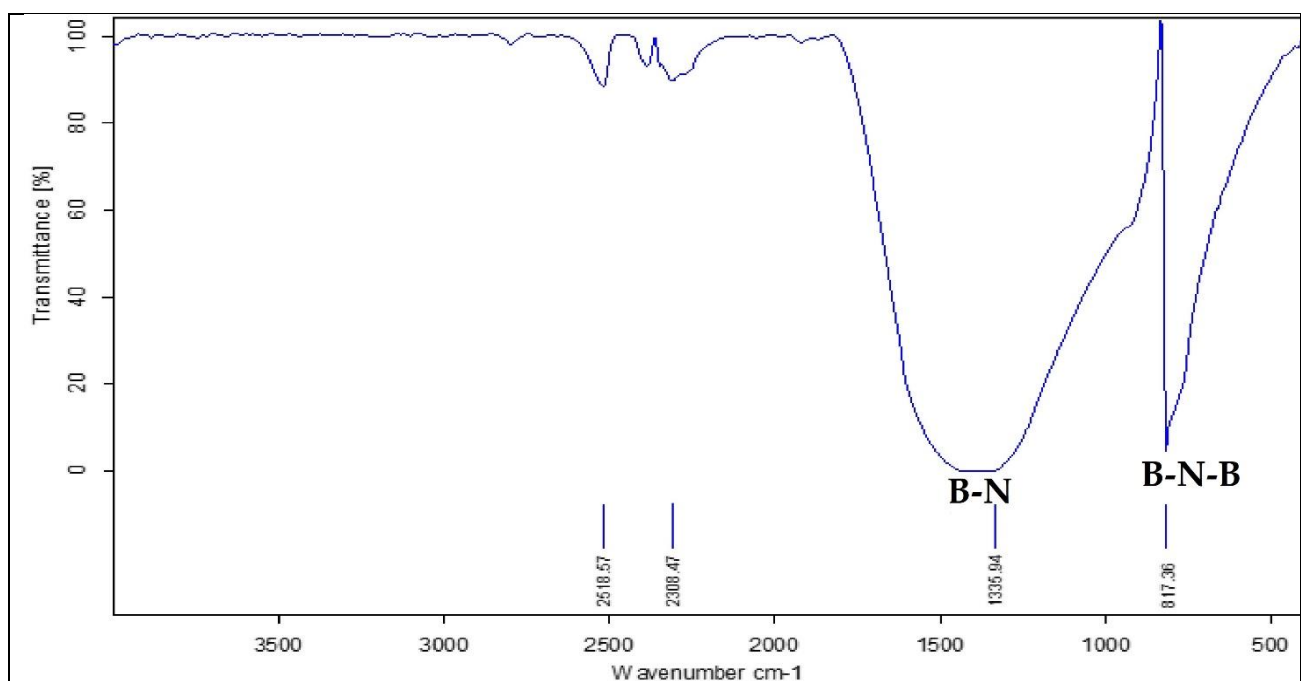

Fig. 6. FTIR spectra of nanoparticles of h-BN of "Chempur" company obtained after two years of exposure to air/moisture and surface gas adsorption.

5. **Conclusion**

This study demonstrates that boron nitride (BN) nanopowders synthesized under high-temperature, non-equilibrium conditions in a high-flux optical furnace as received exhibit strong electrostatic charging, photoluminescence (PL), and paramagnetic activity. Due to incomplete chemical reactions, the surface of the resulting nanoparticles is covered with a thin layer of sassolite, followed by boron oxides, beneath which lies a boron nitride shell. The subsurface contains boron-rich nitride phases, while the core consists of elemental boron.

Over time, exposure to atmospheric oxygen and moisture leads to surface oxidation, hydroxylation, and passivation of defect states—resulting in the complete suppression of electrostatic charging, PL, and electron paramagnetic resonance (EPR) signals. FTIR, PL, and EPR measurements collectively reveal that surface-related defects—rather than bulk properties—predominantly govern the optical and magnetic responses of BN nanostructures.

Therefore, FTIR, PL, and EPR should be regarded as complementary tools for assessing surface evolution and defect dynamics in BN nanopowders, providing valuable guidance for materials stabilization and application-specific tailoring. These findings further emphasize that spontaneous environmental passivation fundamentally differs from intentional doping strategies, highlighting the importance of controlled surface engineering in preserving the functional performance of BN-based nanomaterials.

**ACKNOWLEDGMENTS**

I acknowledge the support of JSPS and ASP NANU. My best acknowledgments for support came from Prof. Hirofumi Takikawa, Prof. Mototsugu Sakai, and Associate Prof. Hiroyuki Muto.

**Declaration of generative AI and AI-assisted technologies in the writing process**

During the preparation of this work, the author Lina L. Sartinska used ChatGPT 1.1.0 to improve her English and for a better understanding of the results of the research. After using this tool/service, Lina L. Sartinska reviewed and edited the content as needed and took full responsibility for the content of the publication.